%% file: samples/MAIN.tex
\documentclass[sigconf]{acmart}

\usepackage{colortbl}
\usepackage{floatflt}
\usepackage{wrapfig}

\definecolor{Gray1}{gray}{0.9}
\definecolor{LightGreen}{rgb}{1,1,0.92}
\definecolor{LightCyan}{rgb}{0.95,1,1}

\AtBeginDocument{%
  }

\setcopyright{acmlicensed}
\copyrightyear{2025}
\acmYear{2025}
\acmDOI{XXXXXXX.XXXXXXX}
\acmConference[WORKS 2025]{}{November 17, 2025}{St. Louis, MO}
\acmISBN{978-1-4503-XXXX-X/2025/11}

\begin{document}

\title{A Study on Messaging Trade-offs in Data
Streaming for Scientific Workflows}
\thanks{This manuscript has been authored by UT-Battelle, LLC under
  Contract No. DE-AC05-00OR22725 with the U.S. Department of
  Energy. The United States Government retains and the publisher, by
  accepting the article for publication, acknowledges that the United
  States Government retains a non-exclusive, paid-up, irrevocable,
  world-wide license to publish or reproduce the published form of
  this manuscript, or allow others to do so, for United States
  Government purposes. The Department of Energy will provide public
  access to these results of federally sponsored research in
  accordance with the DOE Public Access
  Plan. (http://energy.gov/downloads/doe-public-access-plan).}


\author{Anjus George}
\affiliation{%
  \institution{National Center for Computational Sciences, Oak Ridge National Laboratory}
  \city{Oak Ridge, TN}
  \country{USA}}
\email{georgea@ornl.gov}

\author{Michael~J. Brim}
\affiliation{%
  \institution{National Center for Computational Sciences, Oak Ridge National Laboratory}
  \city{Oak Ridge, TN}
  \country{USA}}
\email{brimmj@ornl.gov}

\author{Christopher Zimmer}
\affiliation{%
  \institution{National Center for Computational Sciences, Oak Ridge National Laboratory}
  \city{Oak Ridge, TN}
  \country{USA}}
\email{zimmercj@ornl.gov}

\author{Tyler~J. Skluzacek}
\affiliation{%
  \institution{National Center for Computational Sciences, Oak Ridge National Laboratory}
  \city{Oak Ridge, TN}
  \country{USA}}
\email{skluzacektj@ornl.gov}

\author{A.J. Ruckman}
\affiliation{%
  \institution{National Center for Computational Sciences, Oak Ridge National Laboratory}
  \city{Oak Ridge, TN}
  \country{USA}}
\email{ruckmanaj@ornl.gov}

\author{Gustav~R. Jansen}
\affiliation{%
  \institution{National Center for Computational Sciences, Oak Ridge National Laboratory}
  \city{Oak Ridge, TN}
  \country{USA}}
\email{jansengr@ornl.gov}

\author{Sarp Oral}
\affiliation{%
  \institution{National Center for Computational Sciences, Oak Ridge National Laboratory}
  \city{Oak Ridge, TN}
  \country{USA}}
\email{oralhs@ornl.gov}

\renewcommand{\shortauthors}{George et al.}

\begin{abstract}
Memory-to-memory data streaming is essential for modern scientific workflows that require near real-time data analysis, experimental steering, and informed decision-making during experiment execution. It eliminates the latency bottlenecks associated with file-based transfers to parallel storage, enabling rapid data movement between experimental facilities and HPC systems. These tightly coupled experimental-HPC workflows demand low latency, high throughput, and reliable data delivery to support on-the-fly analysis and timely feedback for experimental control. Off-the-shelf messaging frameworks are increasingly considered viable solutions for enabling such direct memory streaming due to their maturity, broad adoption, and ability to abstract core messaging and reliability functionalities from the application layer. However, effectively meeting the workflows’ requirements depends on utilizing the framework’s capabilities and carefully tuning its configurations. 

In this paper, we present a study that investigates the messaging parameters, and their configuration choices that impact the streaming requirements of two representative scientific workflows. We specifically characterize throughput trade-offs associated with reliable message transmission for these workflows. Our study is conducted through streaming simulations using synthetic workloads derived from the Deleria and LCLS workflows, employing the RabbitMQ messaging framework within the context of the Data Streaming to HPC infrastructure at OLCF. Our simulations reveal several key observations and practical insights that help users understand which configurations best meet the needs of their streaming workloads. 
\end{abstract}





\maketitle

\input{samples/tex/intro}
\input{samples/tex/related_work}
\input{samples/tex/data-streaming}
\input{samples/tex/messaging-params}
\input{samples/tex/simulations}
\input{samples/tex/discussion}

\begin{acks}
This research used resources of the Oak Ridge Leadership Computing Facility located at Oak Ridge National Laboratory, which is supported by the Office of Science of the Department of Energy under contract No. DE-AC05-00OR22725.
\end{acks}



\end{document}

%% file: samples/tex/intro.tex
\section{Introduction}

The emergence of capable artificial intelligence (AI) methods for accelerating scientific discovery, in the form of foundation models and digital twins, has generated a new impetus for advanced workflows \cite{aicoupledhpcworkflows, material_discovery, weather-forecasting, wes-aicoupled, biomolecular}. An important class of workflows that can incorporate these AI methods is the coupling of scientific instruments such as light or neutron sources and observatories with high performance computing (HPC) systems \cite{junqi-neutron}. For HPC facilities, this is driving significant changes in the way workloads access and utilize compute systems. The U.S. Department of Energy (DOE) Integrated Research Infrastructure (IRI) initiative has a mission to reduce the time to insights from experimental facilities. In their 2023 report~\cite{IRI-ABA-Report}, the IRI task force identified several possible templates for interacting with HPC resources. While most simply utilize store-and-forward mechanisms based on file transfers between facilities, a new method identified is direct memory data streaming, in which compute systems at the experimental facility perform memory-to-memory data movement into compute systems at the HPC facility.

Data streaming offers powerful emerging capabilities such as near real-time data analysis, experimental steering, and informed decision-making during experiment execution. However, data streaming into compute allocations creates several new challenges for HPC centers in the areas of security, operational policy, and technical areas related to networking and libraries \cite{ace_tech_report, etz-isc-paper}. The bandwidth and latency requirements of the experimental facilities and the applications running on HPC systems vary greatly \cite{brimstreaming}. In this work, we seek to understand the trade-offs in different communication parameters that are necessary for developing a versatile and useful data streaming architecture. 

Memory-to-memory streaming architectures have become essential for minimizing latency and enabling responsive analysis. At the Oak Ridge Leadership Computing Facility (OLCF), we have developed one such architecture that supports bidirectional memory-based streaming between external experimental instruments and OLCF’s HPC systems. This architecture allows users to provision and deploy commodity off-the-shelf (COTS) streaming frameworks (e.g., RabbitMQ \cite{RabbitMQ}, Redis \cite{Redis}) on demand to support their experiment-driven data flows. The maturity and wider adoption of off-the-shelf messaging frameworks make them attractive choices for enabling direct memory streaming, as they abstract the complexities of core messaging functions such as connection management, buffering, and delivery guarantees away from the application layer, allowing developers to focus on workflow logic rather than low-level data transport mechanisms. These frameworks are designed to handle common challenges in distributed data movement, such as message reliability, flow control, load balancing, and fault tolerance, without requiring users to build these mechanisms from scratch. 

During the development of our Data Streaming to HPC (DS2HPC) infrastructure, we encountered several challenges related to tuning the underlying communication libraries while creating prototype workflows. Two example workflows that highlight the need for low-latency, high-throughput streaming are GRETA/Deleria~\cite{Cromaz2021} and SLAC-LCLS \cite{lcls}. GRETA, using the Deleria workflow software, enables real-time tracking of gamma-ray energy and 3D positioning, while LCLS supports X-ray scattering experiments for molecular analysis, requiring immediate feedback to adjust beamline parameters. Upcoming systems, such as LCLS-II \cite{lcls-2}, impose strict time constraints for real-time data analysis and experiment steering. These workflows primarily demand low latency, high throughput, and reliable streaming. However, achieving reliability is challenging in real-time settings where unpredictable events, such as network disruptions, node failures, or traffic spikes, can jeopardize data fidelity and overall experimental success. For workflows like Deleria, message loss is considered an error at the detector level, making reliability a critical requirement.

To support the reliable and high-throughput transmission required for such time-sensitive workflows, the underlying messaging or data streaming framework must be utilized efficiently. Most messaging frameworks offer a vast array of configurable parameters, and identifying the right combination to balance reliability and performance is a non-trivial task. While real experiments could be conducted to explore these capabilities, they are often costly in both time and machine resources. For instance, neutron beam streaming experiments may require multi-day reservations of expensive instruments, and a typical GRETA experiment can last for up to five days. The optimal configuration often depends on the specific characteristics of the application, including data size, message rate, and tolerance for latency or message loss. Trial and error in real environments is often impractical; thus, simulation becomes a valuable method for systematically exploring these trade-offs.

In this paper, we present a simulation-based study that investigates the messaging parameters and configuration choices that impact the streaming performance of two IRI science workflows: Deleria and LCLS. We specifically characterize the trade-offs between messaging reliability and throughput for these workflows. We begin by identifying the data streaming characteristics of these workflows and the tunable messaging parameters that influence performance. We selected the RabbitMQ messaging framework, provisioned through our DS2HPC architecture, as the candidate for performing streaming simulations. We then simulate\footnote{Our data streaming simulator and configuration files are publicly available here: \url{https://github.com/Ann-Geo/StreamSim/}} various configurations and evaluate their impact on both throughput and reliability, providing insights into how messaging frameworks can be tuned to align with workflow requirements in time-sensitive experimental computing environments. Our simulations identify several key messaging parameters, their configurations, and their impact on messaging throughput for scientific workflows. 

The rest of the paper is organized as follows. Section \ref{sec:rel-work}, discusses existing literature on data streaming in HPC and messaging trade-off studies. Section \ref{sec:data-stream-iri-work} provides a brief overview of the DS2HPC architecture and the IRI science workflows. Section \ref{sec:parameters} describes the messaging parameters used in our studies. Section \ref{sec:simulations} presents the simulation experiments, methodology, and the throughput trade-off insights derived from the study. Finally, Section \ref{sec:discussions} discusses our conclusions and future directions. 

%% file: samples/tex/related_work.tex
\section{Related Work}
\label{sec:rel-work}

\subsection{Data Streaming in HPC}
Data streaming toolkits, architectures, and infrastructures for edge-to-HPC real-time data analysis and experimental steering have been the focus of several recent studies. SciStream \cite{scistream-paper} is a middlebox-based toolkit designed to address infrastructural challenges that hinder memory-to-memory data streaming between scientific environments that lack direct network connectivity. It introduces a suite of protocols to establish authenticated, transparent connections between producers and consumers in different security domains via intermediate gateway nodes. HStream \cite{hstream} is a data streaming engine aimed at supporting the high-throughput demands of scientific applications by separating the data and compute planes, allowing fine-grained control of each, alleviating memory pressure, and avoiding thrashing. 

In \cite{anl-aps}, the authors present streaming data directly from ANL’s Advanced Photon Source (APS) instruments to the Polaris supercomputer using an EPICS-based streaming framework to facilitate real-time analysis and immediate experimental feedback. EPICS \cite{epics, epics-arch} is a set of open-source tools, libraries, and applications for building distributed, soft real-time control systems for scientific instruments such as particle accelerators and telescopes. Vescovi et al. describe a study to identify and capture reusable patterns for computational flows from scientific instruments to computing, data repositories, and other resources---patterns ranging from online data processing to ML training and data cataloging \cite{vescovi}. In \cite{etz-isc-paper}, Etz et al. discuss a transition pathway bridging development testbeds and production HPC environments, examining policy and technological requirements through a case study on enabling the LCLStream real-time experimental workflow for near real-time analysis on OLCF’s ACE infrastructure. Each of these works presents a different aspect and goal of data streaming. The DS2HPC infrastructure at OLCF addresses the challenge of enabling secure, bidirectional, memory-to-memory data streaming between science facilities and computing facilities. 

\subsection{Messaging Trade-off Studies}

Studies on communication and system parameter trade-offs have produced several useful techniques across domains such as cellular networks, wireless sensor networks (WSNs), distributed computing, and edge computing. Soret et al. investigate latency-reliability-throughput trade-offs in cellular networks by identifying key sources of variability and designing an analytical framework to quantify them \cite{celluar-network}. In \cite{wsn}, a cost function for inter-cluster routing path selection in WSNs exploits the trade-off between energy and reliability, achieving improved reliability and energy utilization. In \cite{pmrt}, the authors study the minimum number of communication rounds required by a Perfectly Reliable Message Transmission (PRMT) protocol to send a message within given communication limits, analyzing trade-offs among network connectivity, round complexity, and communication complexity. \cite{dist-system} considers expected loss rate in distributed messaging services as a performability measure, deriving closed-form approximations for different quality-of-service settings and showing how the model can guide design trade-offs. In \cite{edge-comp}, four performance aspects of data stream classification in edge computing, energy consumption, predictive performance, memory cost, and time cost are evaluated for trade-off optimization. 

Our study differs from the above in both domain and focus: we examine data streaming to HPC, investigating messaging parameters on a new edge-to-HPC streaming capability, DS2HPC. We identify configuration choices that impact the streaming requirements of IRI scientific workflows and characterize the trade-offs for reliable message transmission in these workflows. 

%% file: samples/tex/data-streaming.tex
\section{Data Streaming and IRI Workflows}
\label{sec:data-stream-iri-work}

\subsection{Data Streaming to HPC at OLCF}
\begin{figure}[ht]
  \centering
  \vspace*{-4mm}
  \includegraphics[width=\linewidth]{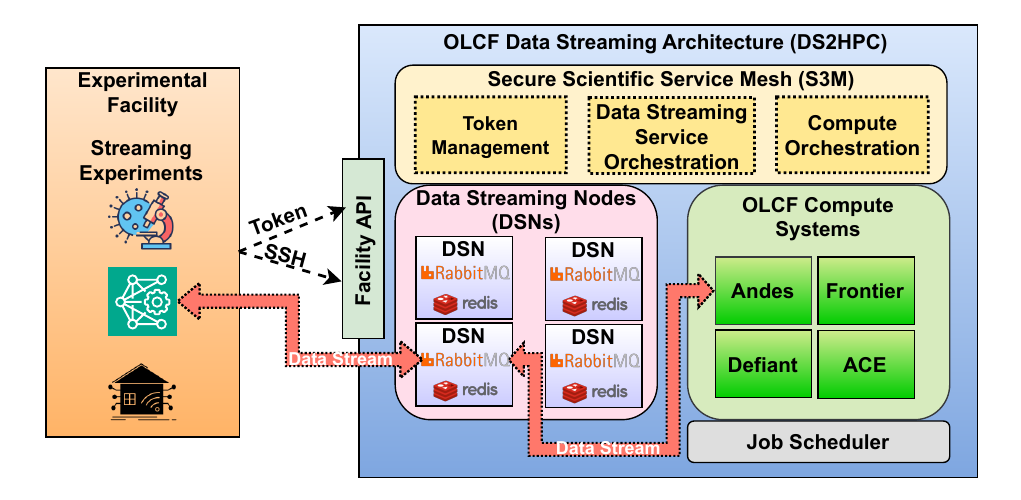}
  \vspace*{-7mm}
  \caption{Data Streaming to HPC (DS2HPC) architecture at OLCF}
  \label{fig:data-stream-s3m}
  \vspace*{-3mm}
\end{figure}

The Oak Ridge Leadership Computing Facility (OLCF) has designed~\cite{brimstreaming} and deployed a new capability for data streaming, called \textit{Data Streaming to HPC (DS2HPC)}, an infrastructure that enables cross-facility scientific workflows through support for bidirectional streaming of data between the memory of producers and consumers distributed between experimental DOE user facilities and OLCF HPC systems (see Figure \ref{fig:data-stream-s3m}). 
Memory-to-memory data exchange between producer and consumer endpoints significantly reduces latency compared to traditional file-based transfers~\cite{abbasi2009datastager}. The decreased latency opens the possibility to incorporate online experiment analysis and feedback applications that may be used for experiment steering, or for rapid generation of follow-on experiment configurations.

Many scientific workflows at the OLCF require tightly coupled orchestration of both compute jobs and data streaming infrastructure. To support these needs, the Secure Scientific Service Mesh (S3M)~\cite{olcf-s3m} serves as OLCF’s primary facility API, offering both RESTful and gRPC interfaces to internal computing resources. S3M is built on a robust security framework that employs time-limited, token-based authentication, enabling external users—including intelligent agents and scientific instruments—to interact securely with OLCF systems. At its core, S3M leverages Istio, a production-grade open-source service mesh, to enforce multi-layered request validation, including authentication, authorization, and fine-grained policy compliance. Each API request is validated against the requester’s project allocations and access permissions using project-scoped tokens issued via a secure web interface.


One of S3M’s key capabilities is the Data Streaming Service Orchestration microservice, which provisions user-specific data streaming services on dedicated Data Streaming Nodes (DSNs) through Kubernetes. DSNs are specialized gateway nodes with high-speed network interfaces connecting both to the public Internet and OLCF’s internal HPC network. Unlike traditional Data Transfer Nodes (DTNs)—which are shared across users and optimized for file-based transfers—DSNs are exclusively allocated to a given workflow, ensuring consistent throughput and minimal latency. These nodes can operate at the application layer (OSI Layer 7), supporting services such as RabbitMQ \cite{RabbitMQ} and Redis \cite{Redis} for pub-sub or message queue patterns, or be configured as Layer 4 routers to forward selected traffic streams directly between systems. By abstracting the complexity of provisioning and securing these services, S3M enables near real-time data exchange for tightly coupled experimental and computational workflows.

\subsubsection{RabbitMQ}
Although both RabbitMQ and Redis clusters can be provisioned using S3M, we use RabbitMQ for this study due to its widespread adoption in previous scientific ecosystem initiatives like INTERSECT \cite{intersect-toolkit, intersect-paper2}. 

\begin{figure}[h!]
\centerline{
\includegraphics[width=0.55\linewidth]{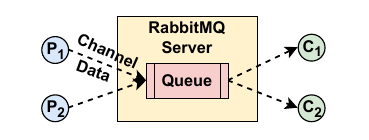}}
\vspace*{-4mm}
\caption{Producers ($P_1$ and $P_2$) and consumers ($C_1$ and $C_2$) using a queue implemented by a RabbitMQ server.}
\label{fig:rabbit}
\vspace*{-0.05mm}
\end{figure}
RabbitMQ is a messaging broker that implements the Advanced Message Queuing Protocol (AMQP) \cite{amqp} as its wire-level communication protocol. A RabbitMQ server (broker) enables clients to send, receive, or temporarily store messages using queues (i.e., ordered collections of messages that are held until consumed). As shown in Figure \ref{fig:rabbit}, a client can act as a producer, which publishes or sends messages, or as a consumer, which receives or consumes messages. A typical operation involves first creating a connection, which abstracts the socket connection and handles protocol negotiation and authentication. Once connected, a channel is opened and all protocol operations including queue declaration and data transfers, occur over this channel. A queue must be declared before sending messages. 

\subsection{IRI Science Workflows}
\label{sec:iri-sci-work}
The OLCF Science Pilots and Workflows initiative aims to implement IRI science workflows on the Advanced Computing Ecosystem (ACE) testbed to enable experimental steering and cross-facility integration across diverse scientific domains. Two representative workflows, GRETA/Deleria and SLAC-LCLS, serve as strong use cases for real-time data streaming. 

GRETA (Gamma-Ray Energy Tracking Array) is a gamma-ray spectrometer currently being deployed at the Facility for Rare Isotope Beams at Michigan State University. It enables real-time analysis of gamma-ray energy and 3D position with up to 100x greater sensitivity than existing detectors. The associated workflow software, Deleria, continuously streams experimental data over ESNet to hundreds of analysis processes on an HPC system, processing up to 500K events per second. Deleria supports time-sensitive streaming and has been deployed across ESNet and the ACE testbed to demonstrate a distributed experimental pipeline. Recent emulation experiments on ACE scaled to 120 simulated detectors, achieving sustained bi-directional streaming rates of $\sim$35 Gb/s. 

The Linac Coherent Light Source (LCLS) at SLAC National Accelerator Laboratory provides X-ray scattering for molecular structure analysis and streams experimental data to enable rapid analysis and decision-making between experiment runs. The LCLStream pilot project trains a generalist AI model using streamed detector data, from both archived and live LCLS/LCLS-II experiments, to support tasks like hit classification, Bragg peak segmentation, and image reconstruction. This AI-driven approach serves as a shared backbone for various downstream data analysis tasks. With the new LCLS-II producing data at 400x the rate of its predecessor, streaming up to 100 GB/s to HPC systems will be essential for responsive analysis and experiment steering. LCLStream aims to support online streaming and real-time analysis during experiment execution, eliminating delays associated with waiting for data to be written to file storage systems before processing. 

Table \ref{tab:workflows} shows the key data streaming characteristics relevant to the two IRI workflows. These characteristics were derived by analyzing aspects such as data size, format, rate, and the nature of producers and consumers involved, as these factors directly influence the streaming behavior. 

The LCLS stream uses 1 MiB data payloads with a steady data rate of $\sim$30 Gbps sustained over 1–100 minutes. Each message contains an HDF5-formatted file, with producers and consumers launched using MPI. Messages are pushed to consumers in a round-robin fashion as they become available in the queue. 

In contrast, Deleria streams messages in the KiB range, each containing multiple experimental events batched together. The number of events per message is variable. Data messages use a compressed binary format, while control messages are encoded in JSON. Depending on the type of experiment, the GRETA detector sustains a steady data rate of up to 32 Gbps once an experiment begins. Producers and consumers do not use MPI. Instead, consumers pull event batches asynchronously from a remote forward buffer, while pushing processed events to a remote event builder. 

\begin{table}[h!]
  \caption{Data streaming characteristics for Deleria and LCLS}
  \label{tab:workflows}
  \vspace*{-3mm}
  \fontsize{7pt}{7pt}\selectfont
  \begin{tabular}{|p{1.0in}|p{0.95in}|p{0.95in}|}
    \hline
    \textbf{Characteristics} & \textbf{Deleria} & \textbf{LCLS} \\
    \hline
    Payload size & $\sim$KiB range & $\sim$1 MiB  \\
    \hline
    Payload format & Binary & HDF5  \\
    \hline
    Payload element & Events & Files \\
    \hline
    Data packaging & Variable number of events per message & One file per message  \\
    \hline
    Data rate & 32 Gbps & 30 Gbps  \\
    \hline
    Data flow & Steady & Steady  \\
    \hline
    Consumption mode & Consumers pull and push messages & Messages pushed to consumers  \\
    \hline
    Consumption parallelism & Parallel consumption (non-MPI) & Parallel consumption (MPI-based) \\
    \hline
    Production parallelism & Parallel production (non-MPI) & Parallel production (MPI-based) \\
    \hline
    Consumer message distribution & Round-robin & Round-robin \\
    \hline
  \end{tabular}
  \vspace*{-3mm}
\end{table}

From discussions with domain scientists, we infer that reliability (no message loss), consistent throughput, and latency tolerance are key aspects when streaming data. For LCLS, low latency and high throughput are both critical to support real-time experimental steering. For Deleria, latency is less critical; however, message loss is unacceptable, as it signals an error at the detector level.  

%% file: samples/tex/messaging-params.tex
\section{Messaging Parameters}
\label{sec:parameters}

We now present the messaging parameters identified to enable workflow streaming based on the characteristics derived in Section~\ref{sec:iri-sci-work}. These workflows additionally require low latency for near real-time data streaming. To achieve streaming that aligns with these characteristics and ensures efficient RabbitMQ operation, some parameters are kept fixed, while others remain tunable to explore trade-offs between throughput and reliability in streaming.

\subsection{Fixed Parameters}

\subsubsection{Messaging Model}

RabbitMQ supports several messaging models, including work queue, publish-subscribe, RPC, broadcasting, and direct routing. Among these, we use the \textbf{work queue model}, where messages are distributed across multiple worker processes - or consumers in our case. By default, RabbitMQ dispatches each message to the next available consumer in sequence, ensuring balanced message distribution over time. This aligns with the workflow requirement for round-robin distribution of payloads (as shown in Table \ref{tab:workflows}).

\subsubsection{Queue Model (Type)}

RabbitMQ offers multiple queue types, including classic, quorum, and stream queues. For our workflow simulations, we use \textbf{classic queues}, which are memory-based queues that retain a fixed number of messages in memory and provide configurable durability (persistence). They do not replicate by default, allowing messages to flow quickly from producer to broker to consumer. Their support for in-memory delivery aligns with the workflows’ primary goal of memory-to-memory streaming. 

In contrast, quorum queues are replicated, and due to their replication mechanism, they exhibit roughly 2x lower throughput than classic queues. All messages are persisted to disk, regardless of delivery mode, and non-durable options are not supported. With large messages (e.g., ${\geq 1MiB}$) and a high number of unacknowledged messages, the cleanup of on-disk segments may lag, resulting in a growing disk footprint. Like quorum queues, stream queues also persist messages to disk. They are suited for scenarios such as large fan-outs (multiple consumers receiving the same message), replay support (allowing consumers to start from any point in the log), and large backlogs, where messages are stored efficiently on disk with minimal memory usage. As with quorum queues, the throughput of stream queues decreases with increasing message size and replication level, due to the additional disk I/O and consensus overhead.

\subsubsection{Queue Length}

When provisioning a RabbitMQ cluster using S3M, users can request up to 32 GiB of RAM. We allocate 80\% of this memory for data payload queues, reserving the remaining 20\% to accommodate additional queues used for workflow simulation management, control messages, and other overhead.

\subsubsection{Queue Overflow Policy}

RabbitMQ supports two overflow policies: ``drop-head'', which discards the oldest messages (FIFO) when the queue is full, and ``reject-publish'', which rejects new incoming messages and applies backpressure to producers. We avoid using drop-head since it may lead to message loss if consumers are slow and haven't yet processed the oldest messages. Instead, we use reject-publish, which allows producers to handle rejected messages and attempt republishing.

\subsubsection{Consumer Delivery Model}

RabbitMQ supports two consumer delivery models: push and pull. We use the push model for both workflows, where messages are automatically delivered to consumers as soon as they arrive in the queue - offering lower latency and higher efficiency. In contrast, the pull model requires consumers to poll the queue for messages, which is particularly inefficient when queues remain empty for extended periods.

\subsection{Tunable Parameters}
\label{sec:tunables}

Reliable streaming in the producer-broker-consumer path can be configured at three levels: the producer, the broker, and the consumer. Accordingly, we tune publisher confirms, message durability, and consumer acknowledgements at each of these levels. Additionally, we explore the impact of two more parameters - prefetch count and multi-queue parallelism on messaging throughput.
Table \ref{tab:tunables} summarizes the tunable parameters and the specific configurations explored for each.

\subsubsection{Publisher Confirms}

Publisher confirms allow producers to track whether messages have been successfully accepted by the broker and delivered to queues. When a producer publishes a message, the server responds with an acknowledgement (ack) if successful, or a negative acknowledgement (nack) if not. This mechanism helps identify messages that may need re-publishing in case of broker failures or network issues. RabbitMQ offers three confirm strategies: asynchronous confirms, where messages are published individually and confirms arrive independently (least reliable but non-blocking); synchronous confirms, where each message is published and waits for confirmation before proceeding (most reliable but blocking); and batch confirms, where a group of messages is published and confirmed together (a balance between reliability and performance, with tunable batch size). 

\subsubsection{Durability}

Each published message can be marked as persistent, ensuring it will be stored on disk and recoverable after a node restart. Separately, RabbitMQ queues themselves can be either durable or transient. Durable queues store their metadata on disk and are restored on node boot, along with any persistent messages they contain. In contrast, transient queues store metadata in memory and are deleted upon restart, along with all messages, regardless of their persistence. Additionally, messages marked as transient are always discarded during recovery, even if stored in a durable queue. Classic queues use an on-disk index to track message locations.

\subsubsection{Consumer Acknowledgements}

When RabbitMQ delivers a message to a consumer, it needs confirmation to determine when the message can be safely discarded. This is handled through message acknowledgements, where the consumer explicitly notifies RabbitMQ that the message has been received and processed. If a consumer disconnects or crashes before sending an ack, RabbitMQ assumes the message was not processed and re-queues it for delivery to another consumer, ensuring no message is lost. 

Messages are pushed to consumers with a delivery tag that uniquely identifies each delivery on a channel. RabbitMQ can consider a message delivered either immediately after it’s sent (in automatic ack mode) or only after receiving an explicit ack (in manual ack mode). Automatic acknowledgements offer higher throughput but risk message loss if the consumer fails mid-processing. In contrast, manual acknowledgements are more reliable and can be batched to reduce network overhead. 

\subsubsection{Prefetch Count}

RabbitMQ delivers messages asynchronously, allowing multiple unacknowledged messages to be ``in flight'' on a channel. The prefetch count sets the maximum number of such unacknowledged messages permitted at a time. Once this limit is reached, RabbitMQ pauses delivery on the channel until at least one message is acknowledged. In general, increasing the prefetch count can improve message delivery rates by keeping consumers busier and reducing idle time.

\subsubsection{Multi-Queue Parallelism}

RabbitMQ limits each queue replica to a single CPU core along its hot code path (message routing, delivery, and acknowledgment handling). Therefore, using multiple queues improves CPU utilization on the server node and enables parallel consumption, resulting in increased throughput. 

\begin{table}
  \caption{Tunable parameters and their configurations}
  \label{tab:tunables}
  \vspace*{-3mm}
  \fontsize{7pt}{7pt}\selectfont
  \begin{tabular}{|p{1.25in}|p{1.7in}|}
    \hline
    \textbf{Tunables} & \textbf{Configurations} \\
    \hline
    Publisher confirms & No confirms, asynchronous confirms (per message), synchronous confirms for per message and batch of messages \\
    \hline
    Durability & Non-persistent and persistent messages \\
    \hline
    Consumer acknowledgements & Automatic, manual with single message and batch of messages \\
    \hline
    Prefetch count & Single and batch of messages \\
    \hline
    Multi-queue parallelism & Single and multiple queues \\
    \hline
  \end{tabular}
  \vspace*{-3mm}
\end{table}

%% file: samples/tex/simulations.tex
\section{Streaming Simulations}
\label{sec:simulations}

\subsection{Methodology and Setup}

To simulate the streaming experiments, we developed a Golang-based application using the amqp091-go (version 1.10.0) \cite{amqp-go} RabbitMQ AMQP client library. The simulator accepts the streaming characteristics of workflows, as listed in Table \ref{tab:workflows}. Additionally, the simulator allows specifying the tunable messaging parameters described in Section \ref{sec:tunables} by providing configuration values for each parameter. For a given message count or test duration, the simulator runs the experiment with the specified number of producers and consumers. Each producer is identical in function and is responsible for generating workload based on the input workload characteristics and sending data to the RabbitMQ server according to the specified messaging parameters. Similarly, each consumer is identical and is designed to receive messages from the RabbitMQ server based on the same set of messaging parameters. 

In addition to the producers and consumers, the simulator includes a coordinator component that serves two primary functions. First, it informs producers and consumers about which queues to use. Second, it collects throughput metrics from individual consumers and reports the aggregate throughput for the entire experiment. Each component, upon startup, handles the initialization of all required queues for the experiment run. The simulator supports launching both MPI-based and non-MPI producers and consumers. 

A three-node RabbitMQ cluster running server version 4.0.5 was deployed on the Data Streaming Nodes (DSNs) to conduct the simulations. Each DSN is equipped with two 32-core 2.70~GHz AMD EPYC 9334 processors and 512~GiB of RAM. However, the RabbitMQ server was configured to use only 12 CPUs and 32~GiB RAM. For the simulator clients, a total of 33 nodes from OLCF’s Andes system \cite{andes} were used: 16 nodes for producers, 16 nodes for consumers, and 1 node for the coordinator. Each Andes node is equipped with two 16-core 3.0~GHz AMD EPYC 7302 processors and 256~GiB of RAM. Andes and the DSNs are connected via a 1~Gbps Ethernet network. 

In the following sections, we present the streaming simulation results for both the Deleria and LCLS workloads across different tunable parameters. For simplicity, we refer to the Deleria workflow as \textit{Dstream} and the LCLS workflow as \textit{Lstream} to indicate that these are simulated representations of the real workflows. Note that Table \ref{tab:workflows} indicates that the Dstream workload’s payload size is in the KiB range, is variable, and streams a variable number of events per message. For the purpose of understanding trade-offs, we fix the payload size to 2~KiB per event and the number of events per message to eight, resulting in a 16~KiB message size.

For all simulations, we measured the aggregate throughput in messages per second from all consumers involved in each experiment. Each data point represents the average of three runs, with each run streaming up to 128K messages. Each test is performed by launching an equal number of producers and consumers to evaluate scaling, with consumers started before producers.

For consistency, most simulations use a default configuration of the tunable parameters to evaluate throughput trade-offs relative to this baseline. In certain sections, we explicitly mention and apply a slightly modified configuration as the new baseline to better illustrate specific trade-offs. In all plots, the corresponding default configuration is indicated using black markers. Unless otherwise stated, the default configuration used in most simulations is as follows:
\begin{itemize}
    \item Publisher confirms: Disabled (No confirms) 
    \item Durability: None 
    \item Consumer acknowledgements: Automatic acknowledgements 
    \item Prefetch count: 1 
    \item Multi-queue parallelism: Single queue 
\end{itemize}

We report results as normalized aggregate throughput relative to the baseline throughput (given by 1 consumer, marked with a red circle in all plots), providing a clear comparison of the performance impacts. To isolate the effect of each parameter, we varied only one tunable parameter at a time, while keeping the others fixed.

\subsection{Publisher Confirms}
\begin{figure}[h!]
  \centering
  \includegraphics[width=\linewidth]{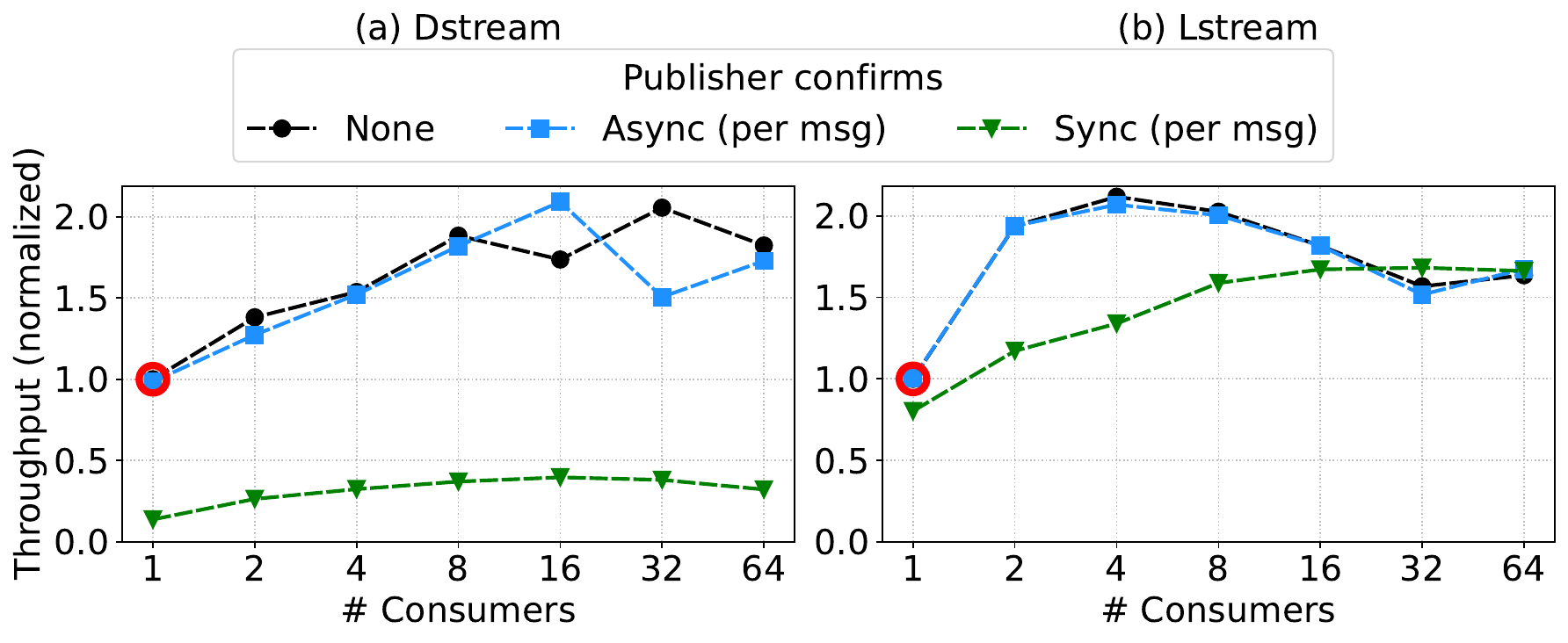}
  \vspace*{-3mm}
  \caption{Throughput for publisher confirm strategies: none, asynchronous per-message confirms, and synchronous per-message confirms for (a) Dstream and (b) Lstream.}
  \label{fig:pub_acks}
  \vspace*{-3mm}
\end{figure}

Figure \ref{fig:pub_acks} shows the throughput trade-offs for different publisher confirm strategies. For both the Dstream and Lstream workloads, the asynchronous per-message confirm mode achieves throughput nearly identical to the least reliable no-confirm mode, with only minimal variation. This is because the publisher does not wait for an acknowledgment from the server before sending the next message. Instead, confirmations can be processed in the background, even after all messages are published, meaning they are not on the critical path of message publishing. 

In contrast, using per-message synchronous confirms results in a significant reduction in throughput - up to 81\% for Dstream and up to 37\% for Lstream. This reduction occurs because the publisher must wait for a server acknowledgment after each message before proceeding. Additionally, we observe that the highest throughput for the Lstream workload is achieved with 4 consumers, after which throughput saturates and converges for higher consumer counts (e.g., 16 and beyond). We attribute this saturation to network bandwidth limitations, as Lstream transmits large payloads (1 MiB per message), which quickly saturate the 1~Gbps Ethernet link between the compute nodes and the RabbitMQ server. 

\begin{figure}[t!]
  \centering
  \includegraphics[width=\linewidth]{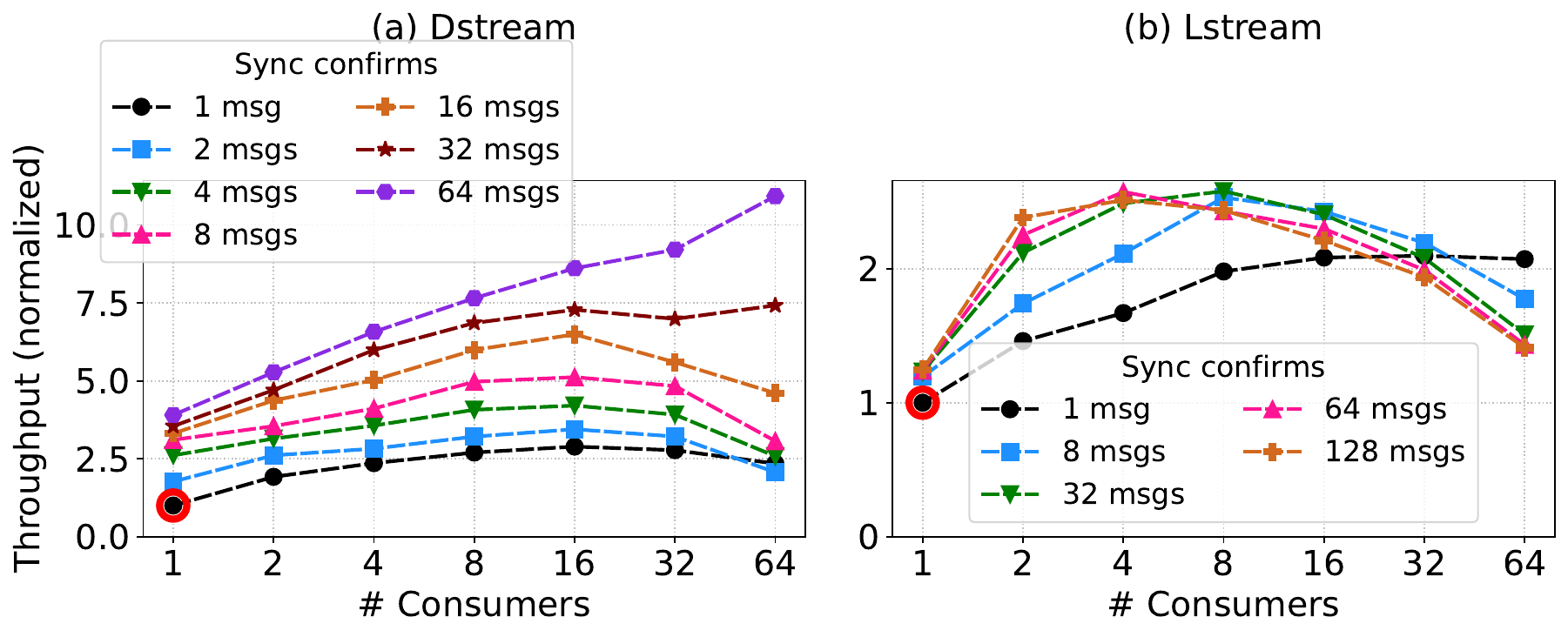}
  \vspace*{-3mm}
  \caption{Throughput for synchronous publisher confirms with different batch sizes for (a) Dstream and (b) Lstream.}
  \label{fig:pub_acks_batch}
  \vspace*{-3mm}
\end{figure}

Figure \ref{fig:pub_acks_batch} shows the throughput trade-offs for synchronous confirms with different batch sizes. In this experiment, the default configuration is set to synchronous per-message confirm (rather than no-confirm) to specifically evaluate how throughput improves when switching to batch-wise confirms. For both workloads, the results demonstrate clear benefits from batching. This is because batching allows the publisher to wait for a single confirmation for many messages, thereby amortizing the confirmation overhead. 
For Dstream, throughput steadily increases as the batch size grows. Specifically, with 64 consumers and a batch size of 64, throughput improves by nearly 4.6x compared to per-message synchronous confirms. For Lstream, the maximum observed improvement is about 50\% with a batch size of 64 in the four consumer case. Further increasing the batch size shows diminishing returns. 

\subsection{Durability}

\begin{figure}[t!]
  \centering
  \includegraphics[width=\linewidth]{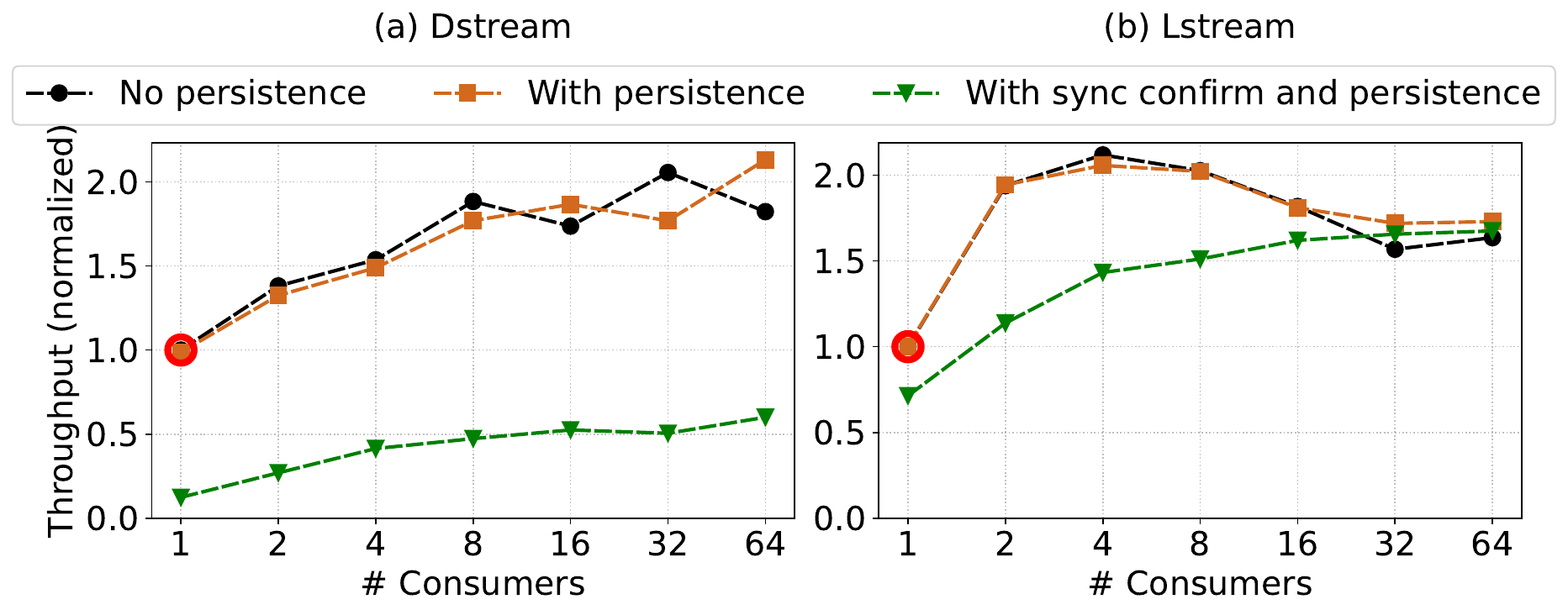}
  \vspace*{-3mm}
  \caption{Throughput for varying message durability: no persistence, with persistence, and with persistence combined with synchronous publisher confirms, for (a) Dstream and (b) Lstream.}
  \label{fig:durability}
  \vspace*{-3mm}
\end{figure}

In Figure \ref{fig:durability}, we observe that the throughput for messages with and without persistence is nearly the same. This is because classic queues in RabbitMQ temporarily store incoming messages in a small in-memory buffer, which is then written to disk either when the buffer fills or based on certain triggers.  Additionally, if messages are consumed quickly, they may not be written to disk at all, as part of RabbitMQ’s performance optimizations. 
However, a persistent message can still be lost if the RabbitMQ node fails before the message is flushed to disk. For example, if a publisher sends a persistent message to a durable queue but does not enable publisher confirms, and the broker crashes before writing the message to disk, that message will be lost. After the broker restarts, the consumer may expect delivery, but it will not happen. To evaluate this scenario, we also measure the throughput for the guaranteed persistence case where publishers use confirms to ensure messages are safely stored before receiving an acknowledgment. In this setup, we observe a throughput reduction of up to 67\% for the Dstream workload, highlighting the performance cost of strict durability guarantees. 

\subsection{Consumer Acknowledgements}

\begin{figure}[ht]
  \centering
  \includegraphics[width=\linewidth]{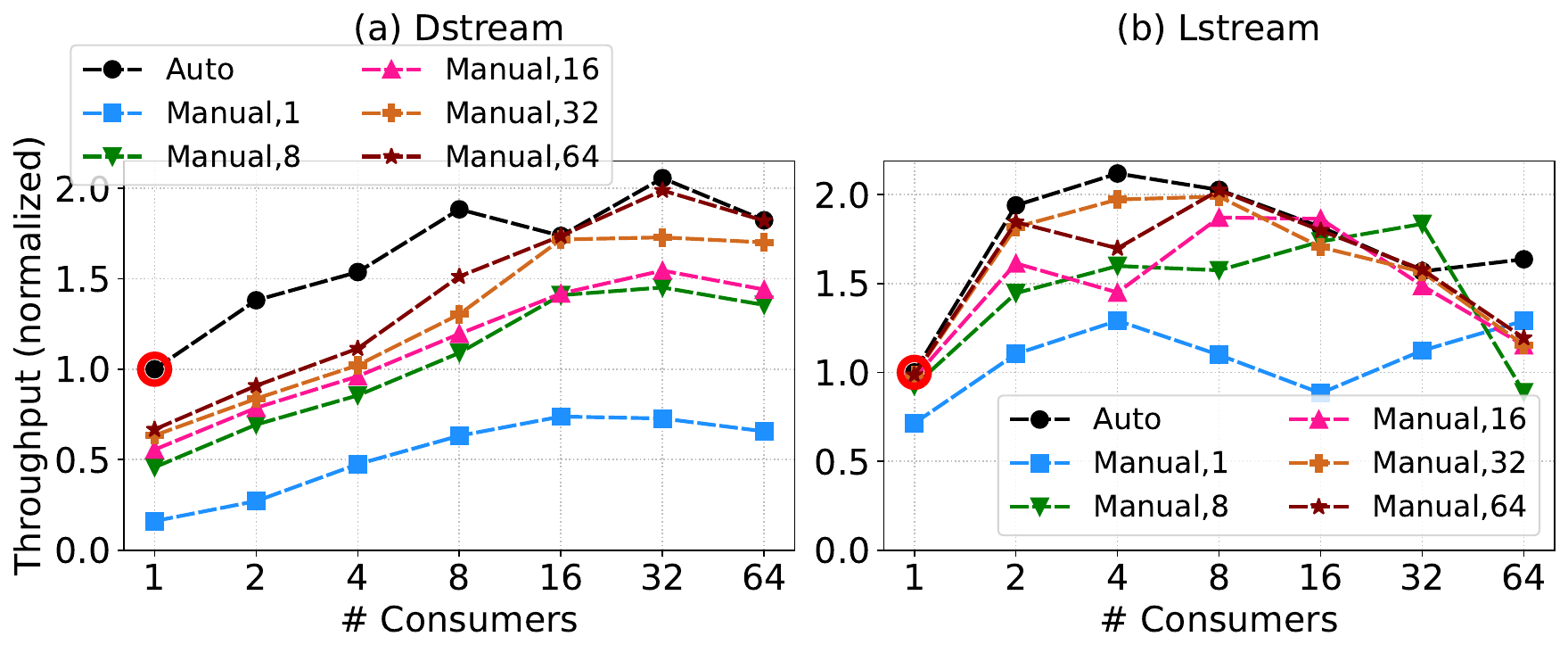}
  \vspace*{-3mm}
  \caption{Throughput for consumer acknowledgment strategies: automatic, manual per message, and manual per batch with varying batch sizes for (a) Dstream and (b) Lstream.}
  \label{fig:cons_acks}
  \vspace*{-3mm}
\end{figure}

Figure \ref{fig:cons_acks} compares the throughput for various consumer acknowledgement strategies. Automatic mode yields the highest throughput because the consumer does not need to send acks back to the server, who assumes messages are successfully received by the consumer. In contrast, the manual per-message acknowledgement mode, which is the most reliable, results in the lowest throughput. 
To enable batch acknowledgements, the consumer must first prefetch the corresponding number of messages, which defines how many messages are allowed to be in flight before sending an acknowledgment. Therefore, for each batch acknowledgement experiment, we adjust the prefetch count from the default value of 1 to match the batch size. The results show that batch acknowledgements significantly improve throughput compared to per-message acknowledgements. For the Dstream workload, acknowledging 64 messages at a time delivers up to 2.7x higher throughput than the per-message ack case and achieves throughput almost equivalent to the auto-ack (default) mode. However, for Lstream, the throughput eventually plateaus, and increasing the batch size beyond 32 yields diminishing returns. This may be due to network I/O limitations or broker memory pressure, caused by a large number of in-flight, unacknowledged messages. 

\subsection{Prefetch Count}

\begin{figure}[t!]
  \centering
  \includegraphics[width=\linewidth]{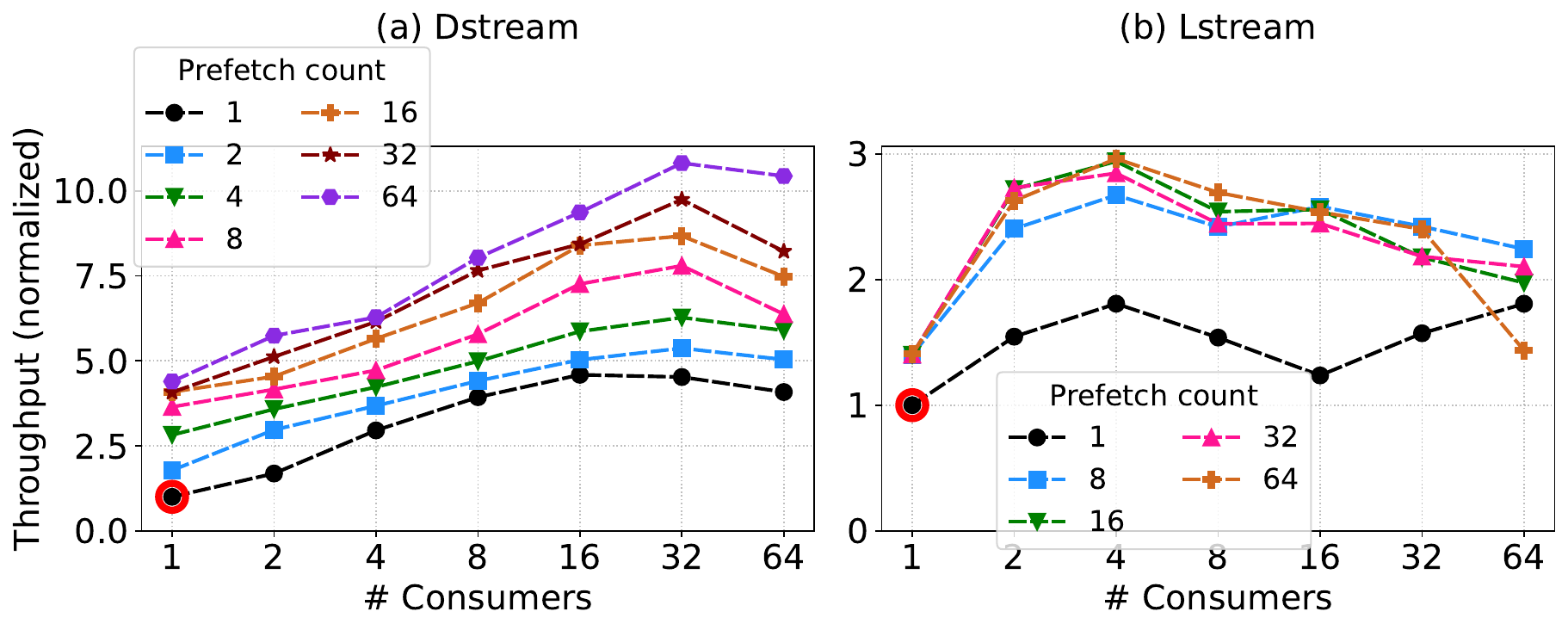}
  \vspace*{-3mm}
  \caption{Throughput for different prefetch counts for (a) Dstream and (b) Lstream.}
  \label{fig:prefetch_count}
  \vspace*{-3mm}
\end{figure}

In the previous experiment, we evaluated the impact of batching consumer acknowledgements, where the prefetch count was set equal to the batch size. In this experiment, we isolate the effect of prefetch count alone by prefetching a batch of messages but sending only a single acknowledgment per message, rather than acknowledging the batch together. The default configuration used here is a prefetch count of 1, with consumers using per-message manual acknowledgements.
As shown in Figure \ref{fig:prefetch_count}, when the prefetch count is set to 1, throughput is the lowest for both workloads. For Dstream, increasing the prefetch count leads to a steady increase in throughput, reaching a maximum at a prefetch count of 64, with an improvement of up to 2.5x. For Lstream, throughput improves by up to 2x under the same conditions. 
However, similar to the previous observation, increasing the prefetch count beyond 64 provides no significant gains for either workload. This plateau occurs for the same reasons discussed earlier, network I/O saturation or memory pressure on the broker due to a high number of in-flight, unacknowledged messages. 

\subsection{Multi-Queue Parallelism}
\begin{figure*}[ht]
  \centering
  \includegraphics[width=\linewidth]{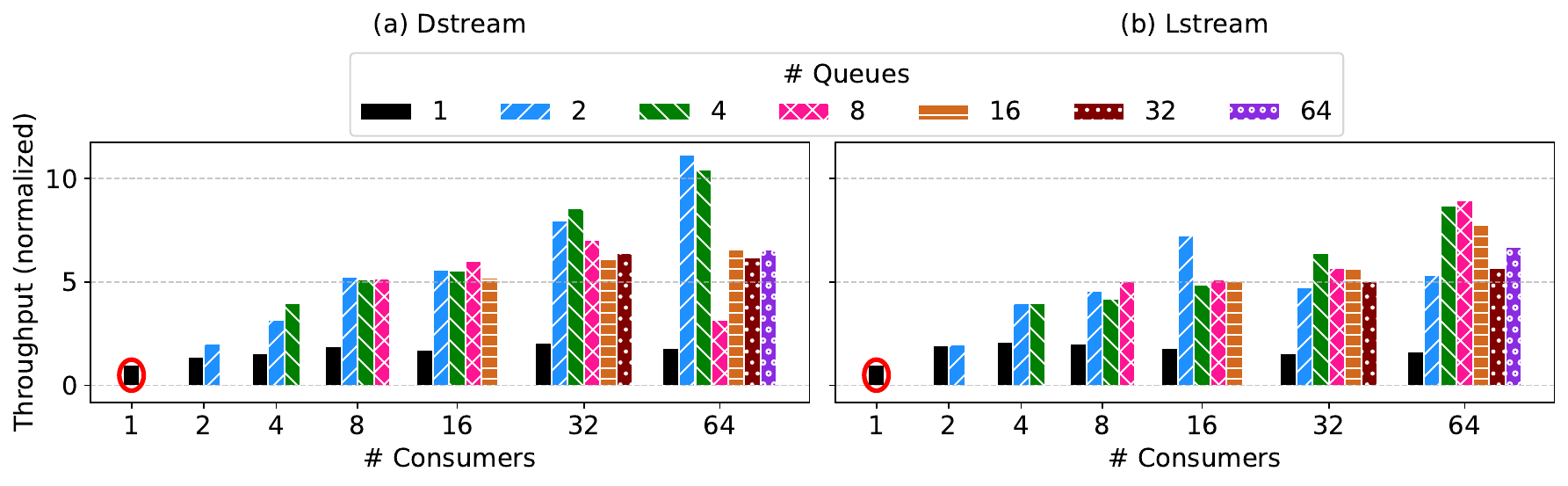}
  \vspace*{-3mm}
  \caption{Throughput for single- vs. multi-queue configurations with parallel consumption for (a) Dstream and (b) Lstream.}
  \label{fig:multi-queue}
  \vspace*{-3mm}
\end{figure*}
To characterize the impact of multi-queue parallelism on throughput, we varied the number of queues from 1 to 64 in powers of 2. In this setup, two consumers share two queues, four consumers are distributed across 1, 2, or 4 queues, and so on. Unlike previous experiments where consumers were started first, this experiment was designed to fully capture the benefits of queue parallelism by pre-filling the queues with messages, meaning producers were started first.

The default configuration for comparison is the scenario where all consumers share a single queue. In Figure \ref{fig:multi-queue}, for the Dstream workload, we observe a clear throughput improvement as the number of queues increases. Notably, when 64 consumers use 2 queues, throughput improves by 6.1x compared to using a single queue. However, increasing the number of queues beyond 4 results in a throughput drop for both the 64- and 32-consumer cases. A similar trend is observed for the Lstream workload, where 16 consumers achieve maximum throughput with 2 queues (a 4x improvement over 1 queue), and 64 consumers achieve peak throughput with 4 queues (a 5.3x improvement over 1 queue). 

Overall, the best throughput occurs when the number of queues is a small fraction of the number of consumers, rather than when the number of queues equals the number of consumers. This is because each RabbitMQ queue runs as an independent Erlang process, and managing a high number of queues introduces overhead from queue state management, memory buffers, and internal flow control. Excessive queues lead to CPU contention and increased scheduler overhead within the RabbitMQ broker. 

%% file: samples/tex/discussion.tex
\section{Discussion and Future Work}
\label{sec:discussions}

In this paper, we presented a study on the throughput trade-offs associated with various messaging parameters and configurations, focusing on balancing reliability, high throughput, and low latency for real-time streaming of scientific workflows. Our study was conducted using the DS2HPC data streaming infrastructure developed at OLCF, which allows users to provision on-demand messaging framework clusters such as RabbitMQ and Redis on high-bandwidth Data Streaming Nodes (DSNs). 

To conduct this study, we selected two representative IRI science workflows and analyzed their data streaming characteristics to derive synthetic streaming workloads that reflect their real-world behaviors. The RabbitMQ-based streaming simulations of these workloads revealed several key insights into how different configurations affect throughput and reliability. Additionally, the results highlight opportunities for further research and exploration in optimizing streaming for scientific workflows. 

\textbf{Choosing optimal configurations}: Our experiments indicate that selecting the optimal messaging configurations for several tunable parameters is not straightforward, as it depends on multiple factors, including the number of producers and consumers, server resources, the inherent design assumptions of the messaging framework, and the interdependencies between these factors. This complexity highlights the value of the simulations we perform, as they help reduce the cost and effort of real experiments by providing users with actionable insights into which configurations can best meet the needs of their workloads. The DS2HPC infrastructure will soon be available to external users, who will be able to access its data streaming capabilities via the S3M API. Our simulations and configuration guidelines provide a reference for such applications when performing streaming on the production-grade infrastructure.

\textbf{Large parameter space}: The experiments we conducted primarily focus on a set of parameters that directly impact throughput, latency, and reliability of message transmission in RabbitMQ. However, RabbitMQ offers a broader range of configurable parameters and settings, including timeouts, routing models, queue types (such as quorum queues and streams), message redelivery policies, and queue properties. While exploring the entire configuration space for generic workloads is practically infeasible, our future work will focus on the targeted selection of configurations for additional science workflows like LCLS and Deleria, where the streaming characteristics are well-defined.

\textbf{Dynamic modeling of workflows}: For the synthetic workloads we derived, we assume that most streaming characteristics remain stable throughout the experiments. For example, in the Deleria workflow, we fixed the number of events per message and payload size to consistently characterize throughput trade-offs, although this value is actually variable in real experiments. Similarly, other factors, such as network congestion, traffic patterns, variable data rates, failure scenarios, and messaging queue backlogs, are inherently more dynamic, particularly because these workflows involve streaming experimental data from external sites or locations. Modeling these dynamic characteristics would provide a more realistic representation of real-world experimental data streaming, and we consider this as another direction for future work.

\textbf{Usage of high-speed networks}: For most experiments with Lstream, it is observed that throughput does not scale as expected beyond 16 consumers. One fundamental reason for this is the use of a 1~Gbps network between the compute nodes (producers/consumers) and the DSNs (RabbitMQ server). The DSNs do contain 100~Gbps network interfaces, but we are still working through issues configuring them to work properly within our OpenShift environment. Once available, we plan to study workloads that use a data-intensive streaming pattern with message sizes in the MiB range while utilizing a 100~Gbps connection to the HPC platform.

\textbf{Choosing the messaging framework}: The current DS2HPC infrastructure supports provisioning of two messaging frameworks - RabbitMQ and Redis. The choice of framework depends on the workflow developer’s preference and the framework’s suitability to the workflow. For this study, we used RabbitMQ, as it has been widely adopted in previous projects such as INTERSECT \cite{intersect-toolkit, intersect-paper2}, where it serves as the message broker and facilitates interaction between instruments and computing resources. In future work, we plan to extend our simulations to Redis and any other messaging frameworks supported by DS2HPC.